\documentclass[aps]{revtex4}

\usepackage{graphicx}
\def\figsiz{3.0in}
\def\figsiz{5.0in}

\def\calS{{\cal S}}

\begin{document}

\title{ Density functional approach to finite temperature nuclear 
properties and the role of a momentum dependent isovector interaction}

\author{ S.J. Lee$^{1}$ and A.Z. Mekjian$^2$}

\affiliation
{$^1$Department of Physics, Kyung Hee University, Yongin, KyungGiDo, Korea}

\affiliation
{$^2$Department of Physics and Astronomy, Rutgers University,
Piscataway, NJ 08854}


\begin{abstract}
Using a density functional approach based on a Skyrme interaction, 
thermodynamic properties of finite nuclei are investigated at non-zero 
temperture. 
The role of a momentum dependent isovector term is now studied besides 
volume, symmetry, surface and Coulomb effects.
Various features associated with both mechanical and chemical instability and
the liquid-gas coexistence curve are sensitive to the Skyrme interaction.
The separated effects of the isoscalar term and the isovector term of momentum 
dependent interaction are studied for a modified SKM($m^*=m$) interaction.
The frequently used Skyrme interaction SLy4 is one of the cases considered
and is shown to have better features for neutron star studies due to
a larger symmetry energy.
\end{abstract}

\pacs
{PACS no.: 24.10.Pa, 21.65.-f, 05.70.-a, 64.10.+h}

\maketitle


\section{Introduction}

  The study of nuclear properties at finite temperature is important 
for nuclear astrophysics, medium energy heavy ion collisions and 
future experiments at the Rare Isotope Beam Facility (RIB). 
It is also of general interest because of its relation to the study of
strongly correlated fermions and to phase transitions in multi-component 
systems, and in particular to these studies in two component systems. 
Density functional theory has been extensively used as an approach 
to such studies in many areas of many body physics. 
In nuclear physics density functional theory based on a Skyrme 
interaction approach has been frequently used. 
Here such an approach will be employed in our study of nuclear 
properties at finite non-zero temperature.
Most applications of density functional theory are
for systems at $T = 0$.

 The nuclear system has terms that set it apart from other two 
component binary systems. 
These include Coulomb terms from the charged proton component, 
symmetry terms from the nuclear symmetry energy, surface energy
terms and momentum dependent 
terms from a velocity dependence of the nuclear interaction. 
Questions related to nuclear stability should involve these terms 
since it is their interplay that determines important features.
For example, fission processes are in part an interplay
of surface versus Coulomb interactions.
The valley of nuclear stability in $N$ and $Z$ involves
an interplay of symmetry versus Coulomb interactions.
While the Coulomb energy favors systems with large neutron excess 
in heavy nuclei, symmetry terms favor systems with equal number of 
protons and neutrons. 
Nuclei far from the valley of nuclear stability up to neutron drip 
and proton drip lines are of interest in future RIB experiments.  
Moreover the Coulomb force is long range and the nuclear force is 
short range and nearly charge independent. 
A charge independent nuclear interaction has an associated 
isospin sysmmetry. The presence of such terms in the interaction 
of nucleons make the nuclear case a unique binary system to study. 
Moreover, realistic nuclei are finite and contain surface terms 
so that phase transitions in finite systems can be studied. 
In this paper we incorporate an additional isovector component 
in the momentum dependence of the interaction, extending our 
previous studies \cite{prc63,plb580,prc77} of properties of heated nuclei. 
In our previous work we considered an isoscalar momentum 
dependent term besides volume, surface, non momentum dependent 
symmetry terms and Coulomb interactions.  
A momentum dependent isovector term and related isovector 
effective mass effects are important in a study of properties 
associated with symmetry terms between protons and neutrons. 
How various terms in nuclear two component systems manifest 
themselves at non-zero temperature and how they affect 
the equation of state, compressibility, chemical and mechanical 
instability and associated liquid gas coexistence curve are 
investigated. Without Coulomb terms, the coexistence curve 
and chemical and mechanical instability curves are symmetric 
about proton fraction $y = 1/2$.
However a stable nucleus such as Pb$^{208}$ at zero temperature 
has proton fraction $y = 82/208 \approx 80/200 = 0.4$
far from being symmetric in proton number 82 and neutron number 126. 
When Coulomb terms are included a large asymmetry appears 
in curves associated with finite temperature properties. 
The symmetry terms partially restores the symmetry of the curves. 
We also extend our previous calculation by considering a 
commonly used Skyrme interaction labeled SLy4 and discuss its properties.

The phase diagram of a one component system is a simple curve 
of pressure versus density determined by a Maxwell construction. 
The phase structure in a binary system is considerably more 
complex than one component systems because of an extra degree 
of freedom associated with the additional component. 
The binodal surface is now determined by pressure, 
temperature and proton fraction. 
The proton fraction can be different in the denser liquid 
phase than the less dense gas phase because of the short 
range of the nuclear symmetry energy. 
In a nuclear two component system, isospin fractionation 
\cite{prl85xu,prl85li,nucl2877,isosp,pr406,anp26}  
is an example where the monomer gas has a large neutron excess.  
Ref.\cite{anp26} is a recent reference that contains further 
references to isospin fractionation. 
The liquid gas phase transition was first treated as 
a one component system \cite{prc27} and later extended to 
two components \cite{prc29} using a Skyrme interaction. 
Ref.\cite{serot} used a relativistic mean field model and 
studied the role of  the symmetry energy on 
the phase structure in detail. 
Inclusion of Coulomb and surface terms can be found 
in Refs.\cite{prc63,plb580,prc77} where large asymmetries 
were shown to appear from the Coulomb term. 
Pawlowski \cite{prc65} has also considered the role of the Coulomb term. 
Some further studies of the nuclear phase transition can be 
found in Refs.\cite{prc67,prl89,npa748,epja25,prl95,sci298,gross,pr389,bertsch,plb650,prc76,prc72}. 
Ref.\cite{plb650} also includes a discussion of momentum 
dependent terms, both isoscalar and isovector. 
Our results will differ somewhat from those of Ref.\cite{plb650}
because we consider a Coulomb term which leads to asymmetries 
in various quantities in proton fraction. 
Without a Coulomb interaction symmetry exists around a proton 
fraction $y = 1/2$ for nuclear stability and for the coexistence loops.
The importance of nuclear isospin symmetry in various features 
of heavy ion collisions can be found in Refs.\cite{isosp,prbali}.

In Sect. II, we give the main equations that are necessary for understanding 
the thermodynamic properties of two component systems with momentum 
dependent interaction at non-zero temperature. 
Sect. III is an application of the results of Sect.II to questions 
associated with the mechanical and chemical instability of hot nuclear 
matter and the associated liquid-gas phase coexistence curve. 
A summary and conclusions are given in Sect. IV.

\section{Themodynaic Properties of Nuclei in a Skyrme Density
Functional Description}

\subsection{General Results}

In this section we present results for the thermodynamic properties
of nuclear matter
which are extended from the results of Ref.\cite{prc63,plb580,prc77} 
to now include a isovector velocity or momentum dependent interaction.
For a Skyrme interaction, this momentum dependent isovector term has some 
important consequences for both nuclear stability and for phase 
transition properties.  

For a nuclear system of proton ($\rho_p$) and neutron ($\rho_n$),
this gives the local potential energy density as
\begin{eqnarray}
 U(\rho_q) &=& \frac{t_0}{2} \left(1 + \frac{x_0}{2}\right) \rho^2
         - \frac{t_0}{2} \left(\frac{1}{2} + x_0\right) \sum_q \rho_q^2
     + \frac{t_3}{12} \left(1 + \frac{x_3}{2}\right) \rho^{\alpha+2}
         - \frac{t_3}{12} \left(\frac{1}{2} + x_3\right)
            \rho^\alpha \sum_q \rho_q^2
              \nonumber   \\
  & & + \frac{1}{4} \left[t_1 \left(1 + \frac{x_1}{2}\right)
         + t_2 \left(1 + \frac{x_2}{2}\right)\right] \rho \tau
      - \frac{1}{4} \left[t_1 \left(\frac{1}{2} + x_1\right)
         - t_2 \left(\frac{1}{2} + x_2\right)\right] \sum_q \rho_q \tau_q
              \nonumber   \\  & &
         + C \rho^\beta \rho_p^2 + C_s \rho^\eta
    \label{potene}
\end{eqnarray}
Here $C \rho^\beta = \frac{4\pi}{5} e^2 R^2$ with $\beta=0$ 
and $C_s \rho^\eta = \frac{4\pi R^2 \sigma(\rho)}{V}
  = \frac{(4\pi r_0^2 \sigma)}{V^{1/3}} \rho^{2/3}$ with $\eta=2/3$ 
when we approximate the Coulomb and surface effects as
coming from a finite uniform sphere of radius $R = r_0 A^{1/3}$
with total charge $Z$ which has the Coulomb energy density
of $U_C = \frac{3}{5}\frac{e^2 Z^2}{R V}$ as discussed in Ref.\cite{prc63}.
We have used the value of $R = 6$ fm and $4\pi r_0^2 \sigma = 20.0$ MeV.
The values for the force parameters used here are given in Table \ref{tabl1}.
We define an effective mass $m_q^*$ as
\begin{eqnarray}
 \frac{m}{m_q^*} &=& 1 + \frac{2m}{\hbar^2} \left\{
        \frac{1}{4} \left[t_1 \left(1 + \frac{x_1}{2}\right)
         + t_2 \left(1 + \frac{x_2}{2}\right)\right] \rho
      - \frac{1}{4} \left[t_1 \left(\frac{1}{2} + x_1\right)
         - t_2 \left(\frac{1}{2} + x_2\right)\right] \rho_q \right\} 
             \nonumber  \\
  &=& 1 + \frac{2m}{\hbar^2} \left\{
        \frac{1}{16} \left[3 t_1 + (5 + 4 x_2) t_2\right] \rho
        \mp \frac{1}{8} \left[t_1 \left(\frac{1}{2} + x_1\right)
         - t_2 \left(\frac{1}{2} + x_2\right)\right] \rho (2y - 1) \right\} 
                \label{effm}
\end{eqnarray}
where $\rho = \rho_p + \rho_n$ and $y = \rho_p/\rho$.
The upper and lower signs are for proton and neutron repectively.
When $x_1 = x_2 = -1/2$, the $m/m_q^*$ is a pure isoscalar
since the coefficient in the term of $\rho_q$ vanishes.
The second form of Eq.(\ref{effm}) shows that the effective masses of 
neutron and proton are the same for a symmetric nuclear system.
We can also see that the neutron effective mass becomes heavier 
and the proton mass becomes lighter in a neutron rich system
while proton effective mass becomes heavier and neutron mass
lighter in a proton rich system.

In Table \ref{tabl1}, the column involving no isovector term
is obtained from the column SKM($m^*=m$) by simply setting
$x_1 = -1/2$ and $x_2 = -1/2$ and the column with a momentum
independent interaction is obtained by setting $t_1 = t_2 = 0$.
Both these columns leave the other parameters the same.
From the table we see that simply changing these parameters
changes various quantities given at the bottom of
the table starting with the effective mass and ending with
the compressibility.
It should be noted that nuclear matter properties for 
the momentum independent interaction 
and the SKM($m^*=m$) are the same except for the symmetry energy.
Specifically, when two interactions have the same effective mass
then the energy and pressure are the same in symmetric systems
with the potential energy of the form of Eq.(\ref{potene})
(see Eqs.(42), (45), and (46) of Ref.\cite{prc77}).
And thus the saturation properties will be the same.
The only difference between these two columns is then the symmetry energy.
By contrast, by simply changing the effective mass, the energy
and pressure in the system are changed and thus also the
saturation properties.
This is seen in Table \ref{tabl1} by comparing the SKM($m^*=m$)
column with the column of no isovector interaction.
Our purpose of just varying $x_1$ and $x_2$ or $t_1$ and $t_2$
without varying other parameters was to isolate the specific
role of the isovector and isoscalar terms in SKM($m^*=m$)
and thus study their role. 
We see that various properties of nuclear matter listed
in the table are changed and the degree to which they are
changed is presented in the results. 
Changing $x_1$ and $x_2$ (or $t_1$ and $t_2$) and also 
changing the other parameters
to fit the saturation properties gives rise to yet another
Skyrme interaction.
Comparisons with many Skyrme interactions can also be made
which we hope to do in the future.

\begin{table}
\caption{Skyrme parameters are in MeV and fm units.
First three columns are from the SKM($m^*=m$) parameter set \cite{prc63,skmm}
except $x_0 = -1/6$ and $x_3 = -1/2$. 
The last column is from the SLy4 parameter set \cite{sly4}.
The nuclear matter properties at the saturation are also shown.
The effective mass is for a saturated symmetric nuclear matter.
  }
  \label{tabl1}
\begin{tabular}{cccccc}
\hline
                        & SKM($m^*=m$) &          &          &   SLy4 \ \ \\
                         &     & \ No Isovector \ & \ Momentum indep. \ &  \\
\hline
     $t_0$               &   --1089.0  & --1089.0 & --1089.0 & --2488.91 \\ 
     $x_0$               &    --1/6    &  --1/6   &  --1/6   &   0.834   \\
     $t_3$               &    17270    &  17270   &  17270   &  13777.0  \\
     $x_3$               &    --1/2    &  --1/2   &  --1/2   &   1.354   \\
   $\alpha$              &      1      &    1     &    1     &    1/6    \\
     $t_1$               &    251.11   &  251.11  &    0     &   486.82  \\
     $x_1$               &      0      &  --1/2   &    0     &  --0.344  \\
     $t_2$               &   --150.66  & --150.66 &    0     & --546.39  \\
     $x_2$               &      0      &  --1/2   &    0     &  --1.000  \\
\ Effective mass $m^*/m$ \ &  0.999987 & 0.894430 &    1     & 0.694658  \\
 Binding energy $E_B/A$  &    15.8173  &  13.3250 & 15.8176  &  15.9722  \\
 Fermi energy   $E_F$    &    34.5186  &  32.0728 & 34.5188  &  36.7743  \\
 Saturation density $\rho_0$ & 0.14509 &  0.12994 & 0.14509  &  0.15954  \\
 Symmetry energy $S_V$   &    18.6080  &  23.7451 & 24.6730  &  32.0038  \\
 Compresibility $\kappa$ &    367.556  &  312.281 & 367.562  &  229.901  \\
\hline
\end{tabular}
\end{table}

For a nuclear system with protons and neutrons with the interaction
given by Eq.(\ref{potene}),
the non-degenerate Fermi gas limit \cite{prc63,plb580,prc77}
leads to the following set of equations.
The chemical potential has a behavior determined by
\begin{eqnarray}
 \mu_q(\rho,y,T) &=& T \ln\left[ \left(\frac{\lambda_q^3}{\gamma}\right)
                               \rho_q \right]
    + \frac{T}{2\sqrt{2}} \left(\frac{\lambda_q^3}{\gamma}\right) \rho_q
           \nonumber \\  & &
   + \frac{1}{4} \left[t_1 \left(1 + \frac{x_1}{2}\right)
       + t_2 \left(1 + \frac{x_2}{2}\right)\right] 
          \frac{3}{2} T \sum_q \frac{2m_q^*}{\hbar^2} 
          \left[\rho_q + \frac{\lambda_q^3}{2^{5/2} \gamma} \rho_q^2\right]
              \nonumber   \\  & &
   - \frac{1}{4} \left[t_1 \left(\frac{1}{2} + x_1\right)
       - t_2 \left(\frac{1}{2} + x_2\right)\right] 
          \frac{3}{2} T \frac{2m_q^*}{\hbar^2} 
          \left[\rho_q + \frac{\lambda_q^3}{2^{5/2} \gamma} \rho_q^2\right]
              \nonumber   \\  & &
   + t_0 \left(1+\frac{x_0}{2}\right) \rho
      + \frac{t_3}{12} \left(1+\frac{x_3}{2}\right) (\alpha+2) \rho^{\alpha+1}
   - \frac{t_3}{12} \left(\frac{1}{2}+x_3\right) \alpha \rho^{\alpha+1}
       \nonumber \\  & &
   - t_0 \left(\frac{1}{2}+x_0\right) \rho_q
 + \frac{t_3}{12} \left(\frac{1}{2}+x_3\right) (\alpha-1) 2 \rho^\alpha \rho_q
 - \frac{t_3}{12} \left(\frac{1}{2}+x_3\right) 2\alpha \rho^{\alpha-1} \rho_q^2
         \nonumber \\  & &
  + C \beta\rho^{\beta-1} \rho_p^2 + 2 C \rho^\beta \rho_p \delta_{q,p}
    + \eta C_s \rho^{\eta-1} .  \label{murqt} 
\end{eqnarray}
The equation of state has a form given by
\begin{eqnarray}
 P(\rho,y,T) &=& \frac{5}{2} T\rho
   + \frac{5}{2} \frac{T}{2\sqrt{2}} \sum_q 
       \left(\frac{\lambda_q^3}{\gamma}\right) \left(\frac{\rho_q^2}{2}\right)
   - \frac{3}{2} T \sum_q \frac{m_q^*}{m} \left[\rho_q
        + \frac{1}{2\sqrt{2}} \left(\frac{\lambda_q^3}{\gamma} \right)
            \left(\frac{\rho_q^2}{2}\right)\right]
                     \nonumber \\  & &
   + \frac{t_0}{2} \left(1 + \frac{x_0}{2}\right) \rho^2
   + \frac{t_3}{12} \left(1 + \frac{x_3}{2}\right)(\alpha+1) \rho^{\alpha+2}
       \nonumber \\  & &
   - \frac{t_0}{2} \left(\frac{1}{2} + x_0\right) \sum_q \rho_q^2
   - \frac{t_3}{12} \left(\frac{1}{2} + x_3\right) (\alpha+1) 
         \rho^\alpha \sum_q \rho_q^2
         \nonumber \\  & &
   + C (\beta + 1) \rho^\beta \rho_p^2  + C_s (\eta - 1) \rho^\eta
               \label{prqt} .  
\end{eqnarray}
The energy density is
\begin{eqnarray}
 {\cal E}(\rho,y,T) &=& \frac{3}{2} T \rho
   + \frac{3}{2} \frac{T}{2\sqrt{2}} \sum_q
     \left(\frac{\lambda_q^3}{\gamma}\right) \left(\frac{\rho_q^2}{2}\right)
                     \nonumber \\  & &
   + \frac{t_0}{2} \left(1 + \frac{x_0}{2}\right) \rho^2
   - \frac{t_0}{2} \left(\frac{1}{2} + x_0\right) \sum_q \rho_q^2
   + \frac{t_3}{12} \left(1 + \frac{x_3}{2}\right) \rho^{\alpha+2}
   - \frac{t_3}{12} \left(\frac{1}{2} + x_3\right) \rho^\alpha \sum_q \rho_q^2
                 \nonumber \\  & &
   + C \rho^\beta \rho_p^2 + C_s \rho^\eta
               \label{erqt}   
\end{eqnarray}
and the entropy is
\begin{eqnarray}
 T\calS(\rho,y,T) &=& \frac{5}{2} T \rho
        - T \sum_q \rho_q \ln\left(\frac{\lambda_q^3}{\gamma} \rho_q\right)
      + \frac{T}{2\sqrt{2}} \sum_q 
        \left(\frac{\lambda_q^3}{\gamma}\right) \left(\frac{\rho_q^2}{4}\right)
               \label{tsrqt}
\end{eqnarray}
The effective mass $m_q^*$ and thus $\lambda_q$ are, in general, isospin
dependent \cite{prc69,prbali}.

For $m/m_q^* = 1 + a_q \rho = 1 + a \rho + b \rho_q 
  = 1 + \rho (a + b [1 \pm (2y-1)]/2)$ 
with $\rho_q = \rho [(1 \mp 1)/2 \pm y]$
and $\lambda_q = \sqrt{2\pi\hbar^2/m_q^*T} = \lambda \sqrt{m/m_q^*}$,
\begin{eqnarray}
 a &=& \frac{m}{2\hbar^2} \left[t_2 \left(1 + \frac{x_2}{2}\right)
              + t_1 \left(1 + \frac{x_1}{2}\right)\right] 
                \\
 b &=& \frac{m}{2\hbar^2} \left[t_2 \left(\frac{1}{2} + x_2\right)
              - t_1 \left(\frac{1}{2} + x_1\right)\right] 
\end{eqnarray}
The $a\rho$ term in the density dependent effective mass comes from 
the isoscalar term of momentum dependent interaction and the $b\rho_q$ term 
comes from the isovector term of momentum dependent interaction.
Using $a$ and $b$, we can study the variation of $P$ and $\mu_q$
with $\rho$ and $y$.
We can study the behavior of thermodynamic quantities at a fixed $P$
using $d P = 0$ from Eq.(\ref{prqt}),
\begin{eqnarray}
 dP &=& \left\{ \sum_q \left(\frac{5}{2} - \frac{3}{2} \frac{m_q^*}{m}\right)
        \left[T \rho_q
         - \frac{T}{2\sqrt{2}} \left(\frac{\lambda_q^3}{\gamma}\right)
              \left(\frac{\rho_q}{2}\right)^2 \right]
                   \right\} \frac{dT}{T}
                     \nonumber \\  
  &+& \left\{\sum_q \left[\left(\frac{5}{2} 
              - \frac{3}{2} \left(\frac{m_q^*}{m}\right)^2\right) T \rho_q
         + \left(\frac{35}{8} - \frac{15}{4} \left(\frac{m_q^*}{m}\right)
                  + \frac{3}{8} \left(\frac{m_q^*}{m}\right)^2 \right)
            \frac{T}{2\sqrt{2}} \left(\frac{\lambda_q^3}{\gamma}\right)
              \left(\rho_q^2\right) \right]  \right.
                     \nonumber \\  & &
   + \left[ {t_0} (1 + \frac{x_0}{2}) \rho^2
   + \frac{t_3}{12} (1 + \frac{x_3}{2})(\alpha+1) (\alpha+2) \rho^{\alpha+2}
          \right.  \nonumber \\  & &
   - {t_0} (\frac{1}{2} + x_0) \left(\sum_q \rho_q^2\right)
   - \frac{t_3}{12} (\frac{1}{2} + x_3) (\alpha + 1) (\alpha + 2)
            \rho^\alpha \left(\sum_q \rho_q^2\right)
         \nonumber \\  & &         \left. \left.
   + C (\beta + 1) (\beta + 2) \rho^\beta \rho_p^2
   + C_s (\eta - 1) \eta \rho^\eta  \right]
                            \right\} \frac{d\rho}{\rho}
           \nonumber \\
   &-& \left\{ \left[ t_0 \left(\frac{1}{2} + x_0\right)
        + \left(\frac{\alpha+1}{6}\right) t_3 \left(\frac{1}{2} + x_3\right)
             \rho^\alpha
        - (\beta+1) C \rho^\beta    
       \right] \rho\rho_3     \right.
       - (\beta + 1) C \rho^{\beta} \rho^2
           \nonumber \\  & &   
     + \sum_q \left[\frac{3}{2} T \rho \left(\pm \frac{m_q^*}{m}\right)
                \left(1 - b \rho_q \left(\frac{m_q^*}{m}\right)\right)
            \right.  \nonumber  \\  & & \hspace{1.0cm}   \left.  \left.
    \mp \frac{T}{2\sqrt{2}} \left(\frac{\lambda_q^3}{\gamma}\right) \rho \rho_q
            \left( \left(\frac{5}{2} - \frac{3}{2} \frac{m_q^*}{m}\right) 
          + b \rho_q \left(\frac{m_q^*}{m}\right) 
            \frac{3}{4} \left(\frac{5}{2} - \frac{1}{2} \frac{m_q^*}{m}\right)
                  \right) \right]
             \right\} dy
\end{eqnarray}
This equation gives the condition determining $y_m(\rho)$ where 
both $\partial P/\partial y = 0$ and $\partial\rho/\partial y = 0$, 
\begin{eqnarray}
 0 &=& 
 \left[ t_0 \left(\frac{1}{2} + x_0\right)
        + \left(\frac{\alpha+1}{6}\right) t_3 \left(\frac{1}{2} + x_3\right)
             \rho^\alpha
        - (\beta+1) C \rho^\beta \right] \rho\rho_3   
       - (\beta + 1) C \rho^{\beta} \rho^2
           \nonumber \\  & &   
     + \sum_q \left[\frac{3}{2} T \rho \left(\pm \frac{m_q^*}{m}\right)
                \left(1 - b \rho_q \left(\frac{m_q^*}{m}\right)\right)
            \right.  \nonumber  \\  & & \hspace{1.0cm}    \left.
    \mp \frac{T}{2\sqrt{2}} \left(\frac{\lambda_q^3}{\gamma}\right) \rho \rho_q
            \left( \left(\frac{5}{2} - \frac{3}{2} \frac{m_q^*}{m}\right)
          + b \rho_q \left(\frac{m_q^*}{m}\right)
            \frac{3}{4} \left(\frac{5}{2} - \frac{1}{2} \frac{m_q^*}{m}\right)
                  \right) \right]    \label{ymin}
\end{eqnarray}
The $y_m(\rho)$ is the value of $y$ with lowest pressure $P$ 
for a given density $\rho$ and temperature $T$
and is indepenent of $\rho$ for a momentum independent Skyrme
interaction with $x_3 = -1/2$ and $\beta = 0$
as considered in Ref.\cite{prc63,plb580}.
The $x_3$ term and the density dependent effective mass for a momentum 
dependent Skyrm force introduce a small $\rho$-dependence in $y_m$.
The isovector momentum dependent term ($x_1$ and $x_2$ terms) introduce
large $\rho$-dependence in $y_m$ as we will see later.
The $y_m(\rho)$ curve cross the coexistence loop at the equal concentration
point $y_E$ where the liguid and gas phases have the same $y$ value.

From Eq.(\ref{murqt}),
\begin{eqnarray}
 d\mu_q &=& \left\{ \ln\left[ \left(\frac{\lambda_q^3}{\gamma}\right)
                           \rho_q \right] - \frac{3}{2}
    - \frac{1}{2} \frac{1}{2\sqrt{2}} 
          \left(\frac{\lambda_q^3}{\gamma}\right) \rho_q
           \right.    \nonumber \\  & &
   + \frac{1}{4} \left[t_1 \left(1 + \frac{x_1}{2}\right)
       + t_2 \left(1 + \frac{x_2}{2}\right)\right] 
          \frac{3}{2} \sum_q \frac{2m_q^*}{\hbar^2} \left[\rho_q 
             - \frac{1}{2} \frac{\lambda_q^3}{2^{5/2} \gamma} \rho_q^2\right]
              \nonumber   \\  & &    \left.
   - \frac{1}{4} \left[t_1 \left(\frac{1}{2} + x_1\right)
       - t_2 \left(\frac{1}{2} + x_2\right)\right] 
          \frac{3}{2} \frac{2m_q^*}{\hbar^2} \left[\rho_q 
             - \frac{1}{2} \frac{\lambda_q^3}{2^{5/2} \gamma} \rho_q^2\right]
          \right\} d T     \nonumber   \\ 
  &+& \left\{ \left[\frac{T}{\rho} 
                + \frac{T}{2\sqrt{2}} \frac{\rho_q}{\rho} 
                   \left(\frac{\lambda_q^3}{\gamma}\right)\right] 
              \left(\frac{5}{2} - \frac{3}{2}\frac{m_q^*}{m}\right) 
           \right.    \nonumber   \\  & &
   + \frac{1}{4} \left[t_1 \left(1 + \frac{x_1}{2}\right)
       + t_2 \left(1 + \frac{x_2}{2}\right)\right] 
          \frac{3}{2} T \sum_q \frac{2m_q^*}{\hbar^2} \frac{\rho_q}{\rho} 
          \left[\frac{m_q^*}{m} 
             + \frac{\lambda_q^3}{2^{5/2} \gamma} \rho_q 
               \left(\frac{5}{2} - \frac{1}{2} \frac{m_q^*}{m}\right)\right]
              \nonumber   \\  & &   
   - \frac{1}{4} \left[t_1 \left(\frac{1}{2} + x_1\right)
       - t_2 \left(\frac{1}{2} + x_2\right)\right] 
          \frac{3}{2} T \frac{2m_q^*}{\hbar^2} \frac{\rho_q}{\rho} 
          \left[\frac{m_q^*}{m} 
             + \frac{\lambda_q^3}{2^{5/2} \gamma} \rho_q 
               \left(\frac{5}{2} - \frac{1}{2} \frac{m_q^*}{m}\right)\right]
             \nonumber \\  & &
   + \left[ t_0 \left(1+\frac{x_0}{2}\right) 
      + \frac{t_3}{12} \left(1+\frac{x_3}{2}\right) 
             (\alpha+2)(\alpha+1) \rho^{\alpha}
   - \frac{t_3}{12} \left(\frac{1}{2}+x_3\right) \alpha(\alpha+1) \rho^{\alpha}
             \right.   \nonumber \\  & &
   - t_0 \left(\frac{1}{2}+x_0\right) \frac{\rho_q}{\rho}
   + \frac{t_3}{12} \left(\frac{1}{2}+x_3\right) 2(\alpha-1)(\alpha+1) 
             \rho^{\alpha-1} \rho_q
   - \frac{t_3}{12} \left(\frac{1}{2}+x_3\right) 2\alpha(\alpha+1) 
             \rho^{\alpha-2} \rho_q^2
         \nonumber \\  & &    \left.   \left.
   + C \beta(\beta+1) \rho^{\beta-2} \rho_p^2 
   + 2 C (\beta+1) \rho^{\beta-1} \rho_p \delta_{q,p}
   + C_s \eta(\eta-1) \rho^{\eta-2} \right]  \right\} d\rho 
         \nonumber   \\
  &+& \left\{ \pm \left[T \frac{\rho}{\rho_q}
                + \frac{T}{2\sqrt{2}} \rho 
                   \left(\frac{\lambda_q^3}{\gamma}\right)\right] 
              \left(1 + \frac{3}{2} b \rho_q \frac{m_q^*}{m}\right) 
           \right.    \nonumber   \\  & &     \hspace{-1.0cm}
   + \frac{1}{4} \left[t_1 \left(1 + \frac{x_1}{2}\right)
       + t_2 \left(1 + \frac{x_2}{2}\right)\right] 
          \frac{3}{2} T \sum_q \left(\pm\right) \frac{2m_q^*}{\hbar^2} \rho 
          \left[\left(1 - b \rho_q \left(\frac{m_q^*}{m}\right)\right) 
             + \frac{\lambda_q^3}{2^{5/2} \gamma} \rho_q 
               \left(2 + b \rho_q \frac{1}{2} \frac{m_q^*}{m}\right)\right]
              \nonumber   \\  & &   
   \mp \frac{1}{4} \left[t_1 \left(\frac{1}{2} + x_1\right)
       - t_2 \left(\frac{1}{2} + x_2\right)\right] 
          \frac{3}{2} T \frac{2m_q^*}{\hbar^2} \rho 
          \left[\left(1 - b \rho_q \left(\frac{m_q^*}{m}\right)\right) 
             + \frac{\lambda_q^3}{2^{5/2} \gamma} \rho_q 
               \left(2 + b \rho_q \frac{1}{2} \frac{m_q^*}{m}\right)\right]
             \nonumber \\  & &
   \pm \left[ - t_0 \left(\frac{1}{2}+x_0\right) \rho
     + \frac{t_3}{12} \left(\frac{1}{2}+x_3\right) 2(\alpha-1)
             \rho^{\alpha+1} 
     - \frac{t_3}{12} \left(\frac{1}{2}+x_3\right) 4\alpha 
             \rho^{\alpha} \rho_q  \right]
         \nonumber \\  & &    \left.   
     + \left[2 C \beta \rho^{\beta} \rho_p 
           + 2 C \rho^{\beta+1} \delta_{q,p} \right] 
       \right\} d y   \label{dmurqt}       
\end{eqnarray}
This equation determines the various curves of $\partial\mu_q/\partial\rho = 0$,
$\partial\mu_q/\partial y = 0$, chemical instability curve, etc.

\section{Applications to Nuclear Mechanical and Chemical Instability
and the Liquid-Gas Phase Transition}  \label{rsltsec}

\subsection{Equation of State with Isoscalar and Isovector
Momentum Dependent Terms}

\begin{figure}
\includegraphics[width=\figsiz]{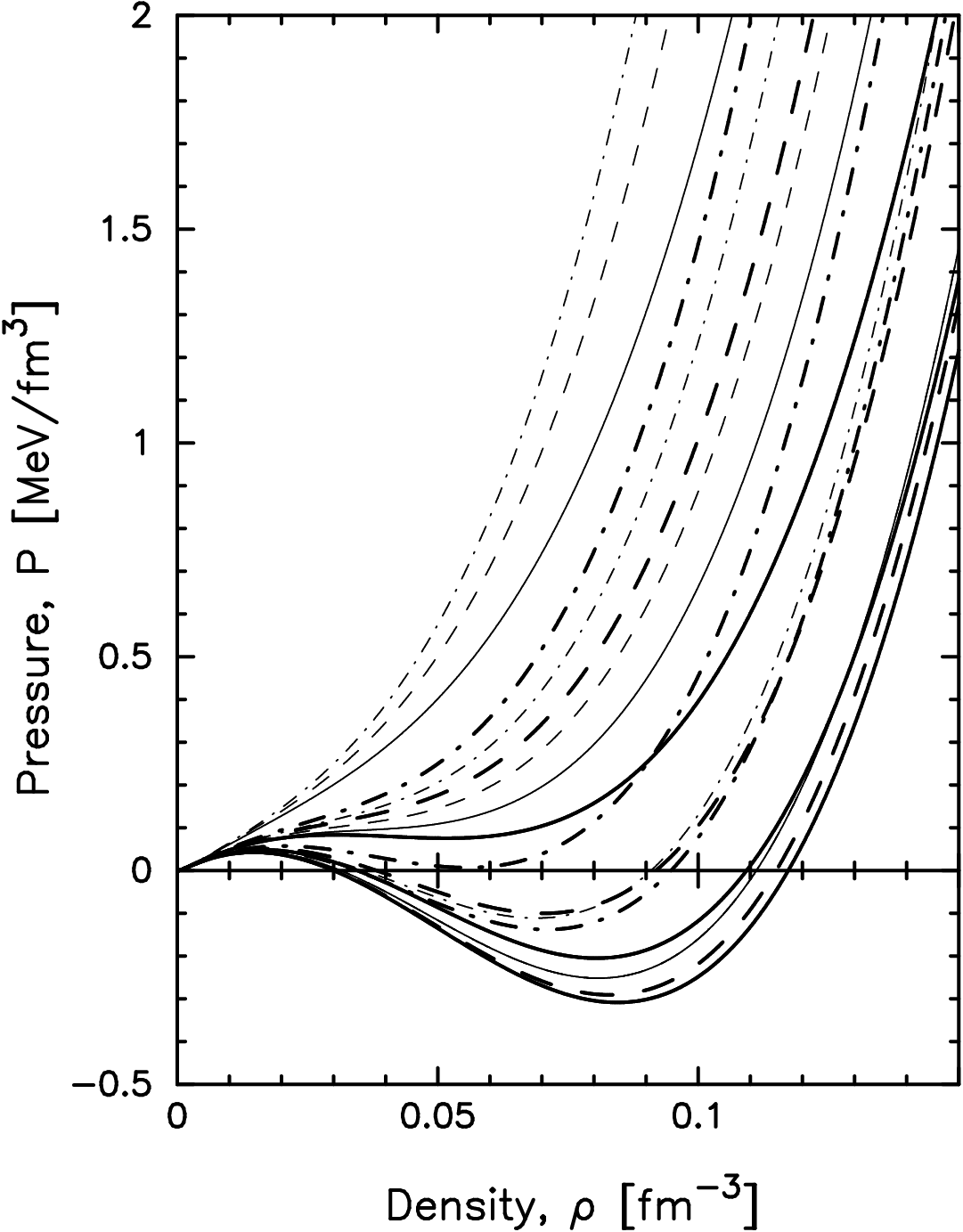}
\caption{Equation of state $P(\rho)$ for fixed $y = 0$, 0.2, $y_m(\rho)$ 
(the proton concentration with minimum pressure defined by Eq.(\ref{ymin})) 
for thick lines from top to bottom
and $y = 0.5$, 0.8, 1.0 for thin lines from bottom to top. 
The solid line is for the modified SKM($m^*=m$) interaction
with both isoscalar and isovector momentum dependent terms, 
the dash-dotted line is for a momentum dependent but
isovector independent term, and 
the dashed line is for momentum independent terms only. 
The lowest thin dashed curve overlaps here with
the lowest thin solid curve which have $y = 0.5$. } 
     \label{fig1}
\end{figure}

\begin{figure}
\includegraphics[width=\figsiz]{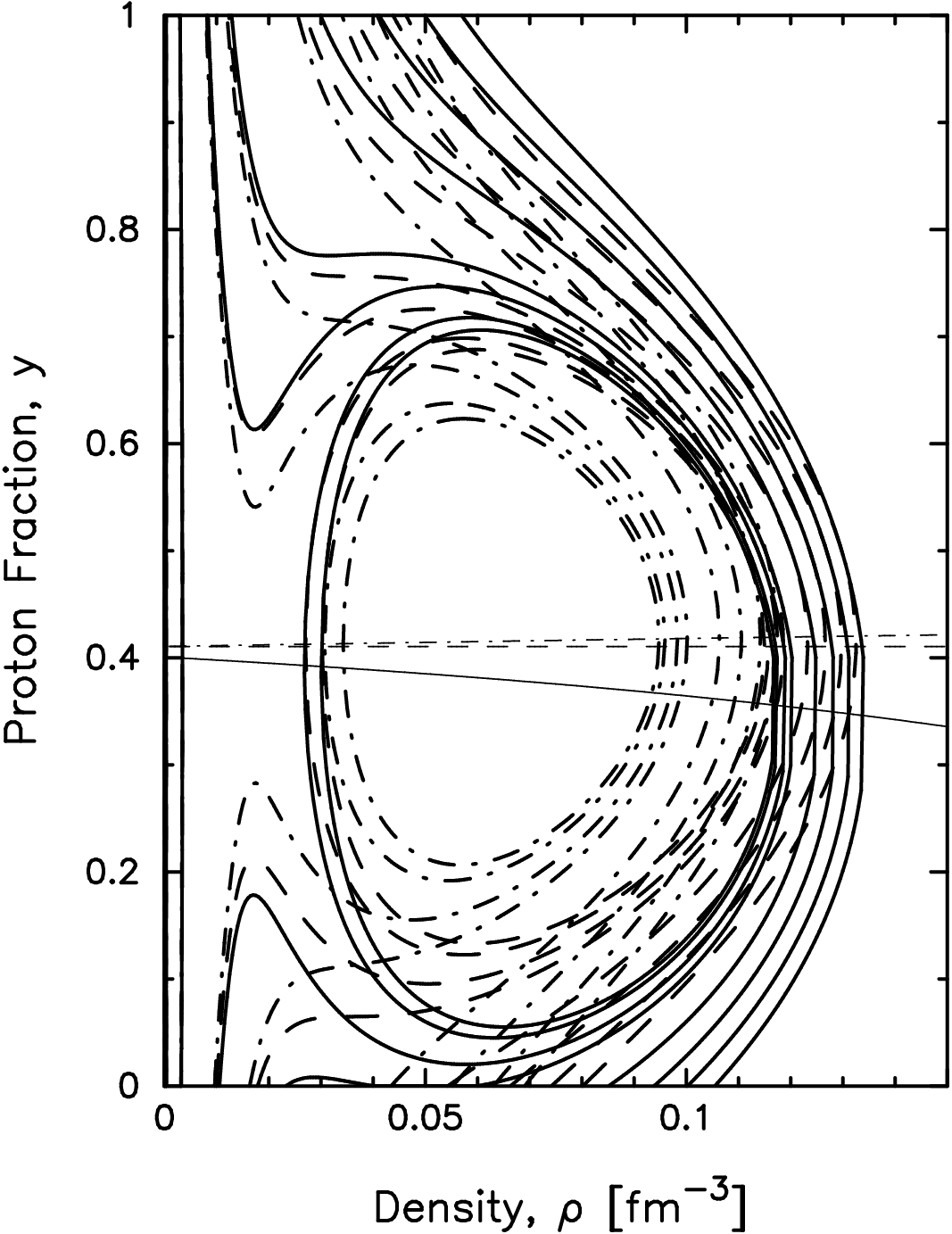}
\caption{The $y(\rho)$ dependence for fixed pressure of $P = 0$, 0.015, 
0.05, 0.08, 0.2, 0.3, 0.4, 0.5 MeV$\cdot$fm$^{-3}$. 
The curves are the same as in Fig.1.
The thin horizontal lines are for $y_m(\rho)$ curve define 
by Eq.(\ref{ymin}). }
     \label{fig2}
\end{figure}

\begin{figure}
\includegraphics[width=\figsiz]{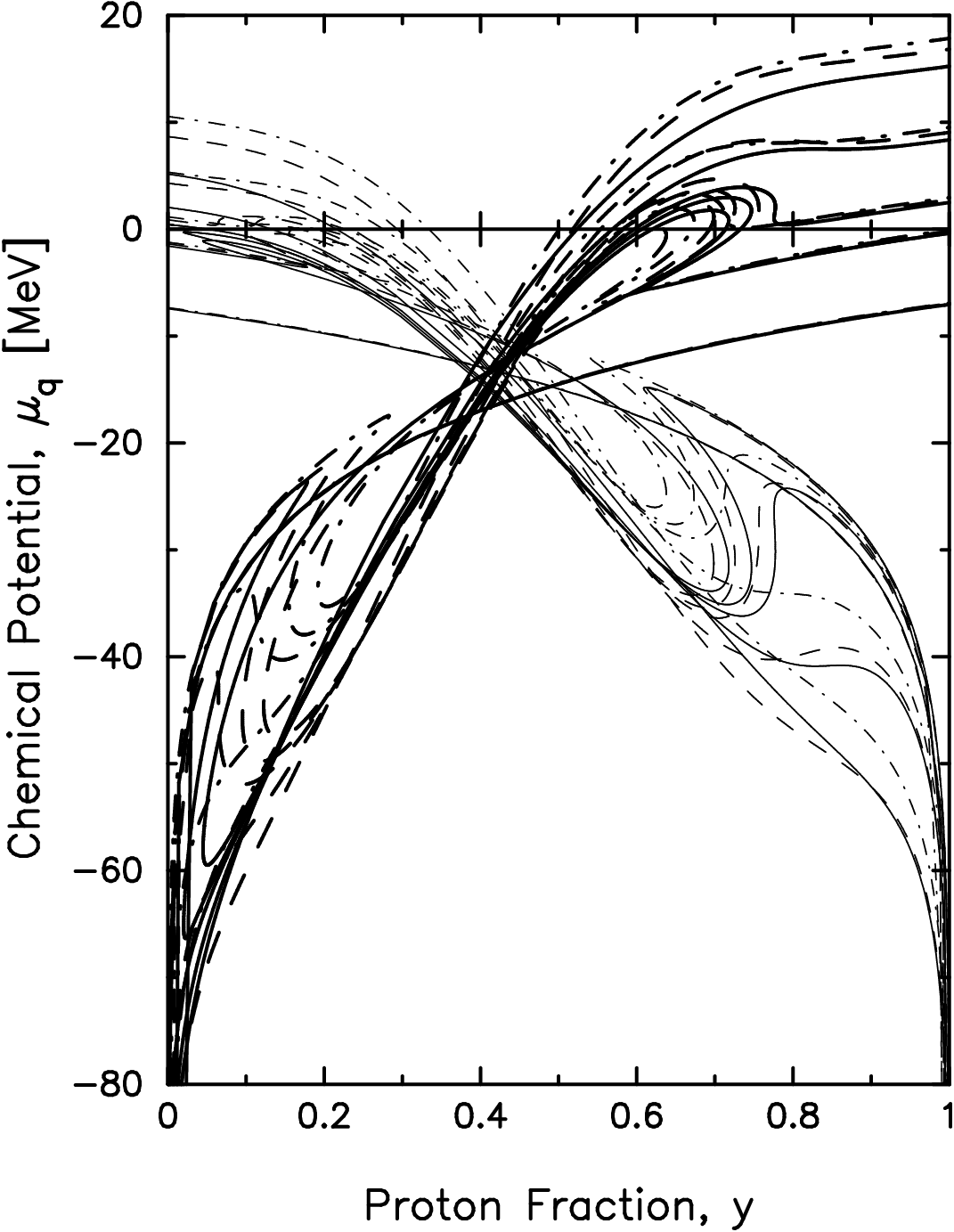}
\caption{Chemical potential $\mu_p$ (thick lines) and $\mu_n$ (thin lines) 
as a function of $y$ for $P = 0.015$, 0.05, 0.08, 0.2, and 0.5.
The curves are the same as in Fig.1 }
     \label{fig3}
\end{figure}

\begin{figure}
\includegraphics[width=\figsiz]{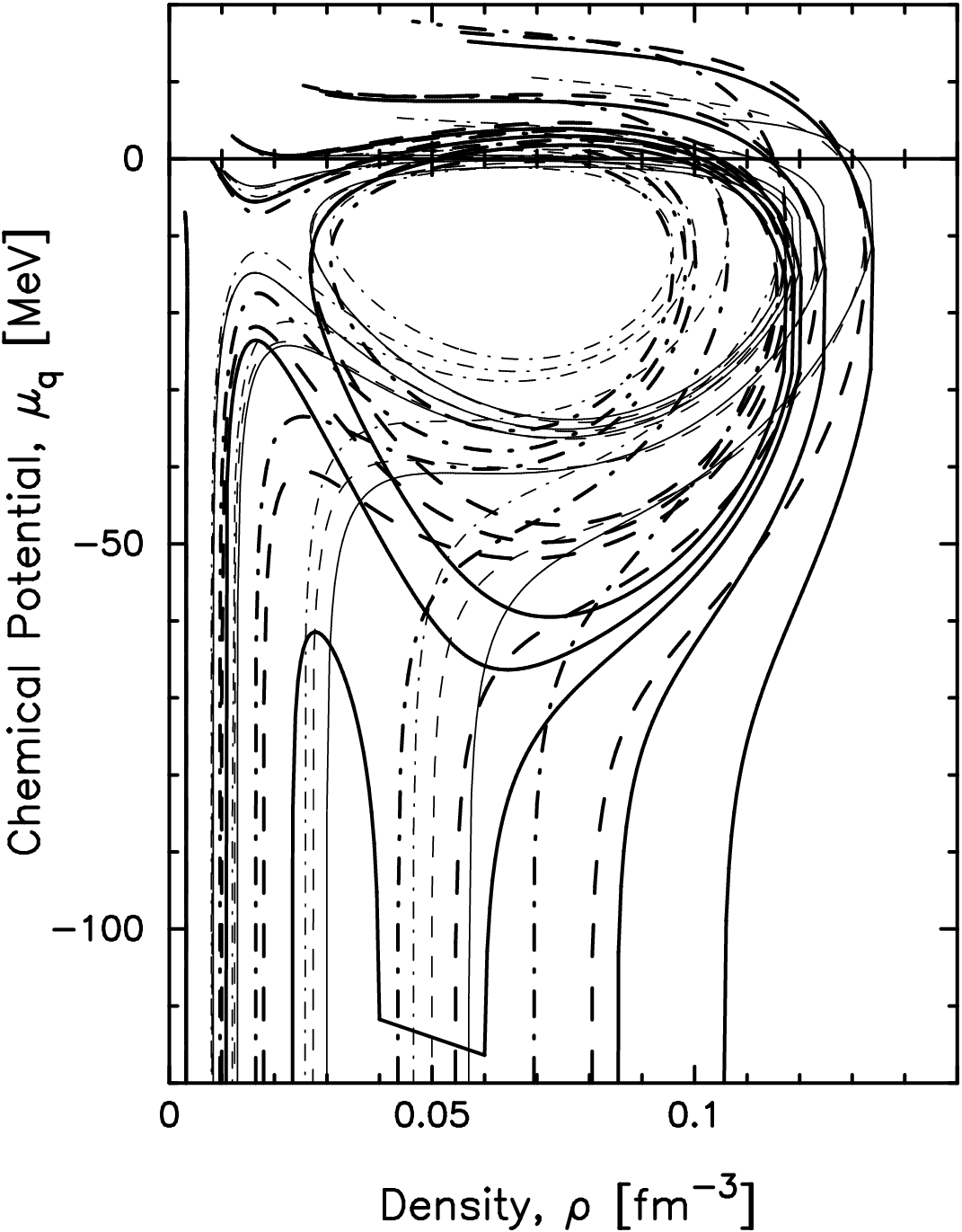}
\caption{Same as Fig.3 but as a function of $\rho$. 
The straight horizontal portion of the thick solid curve (momentum dependent 
force with both isoscalar and isovector terms) for $P = 0.08$ MeV/fm$^3$ 
at low $\mu_p$ just represents the cut 
in the density range of $0.04 < \rho < 0.06$ fm$^{-3}$ 
due to the condition of $y \ge 0$. 
  }
     \label{fig4}
\end{figure}

The equation of states ($P(\rho)$ curve) for various proton fraction $y$ 
at a fixed temperature of $T = 10$ MeV are shown in Fig.\ref{fig1}.
The solid lines are $P(\rho)$ for momentum dependent interaction 
(modified SKM($m^*=m$)) with both isoscalar and isovector terms.
The dash-dotted lines are for a momentum dependent Skyrme interaction 
with isoscalar term only and the dashed lines are 
for a Skyrme interaction without momentum dependent terms 
which are the same cases considered in Ref.\cite{prc77}. 
For various values of pressure $P$ at $T = 10$ MeV, 
the curve of $y(\rho)$, $\mu_q(y)$, and $\mu_q(\rho)$ are shown 
in Fig.\ref{fig2}, Fig.\ref{fig3}, and Fig.\ref{fig4} respectively. 
The basic behaviors of the curves are the same independent of
which force parameters are used.
For example, the $y(\rho)$ and $\mu_q(\rho)$ curves of Figs.\ref{fig2} 
and \ref{fig4} for low pressure $P$ exhibit a closed loop separated from 
a vertical line at low density for all three cases. 
Since the overall behaviors which are indepenent of force parameters 
are studied in previous papers \cite{prc63,plb580,prc77}, 
the discussions below are concentrated more on the differences 
between different force parameters. 

The pressure for a given density $\rho$ and $y$ increases from the 
corresponding value for the case of a momentum independent Skyrme 
interaction (dashed curve) 
by the momentum dependent isoscalar term (dash-dotted curve)
and decreases much more by the isovector term (solid curve).
In Fig.1, the crossing of curves for different $y$ values appears explicitely 
for $y = 0.2$ and 0.5 with a momentum dependent isovector term (solid lines). 
This crossing is related to the $\rho$ dependence of $y_m$, 
the proton concentration with minimum pressure defined by Eq.(\ref{ymin}), 
which comes from the momentum dependent terms and the three 
body isovector term. 

\begin{figure}
\includegraphics[width=\figsiz]{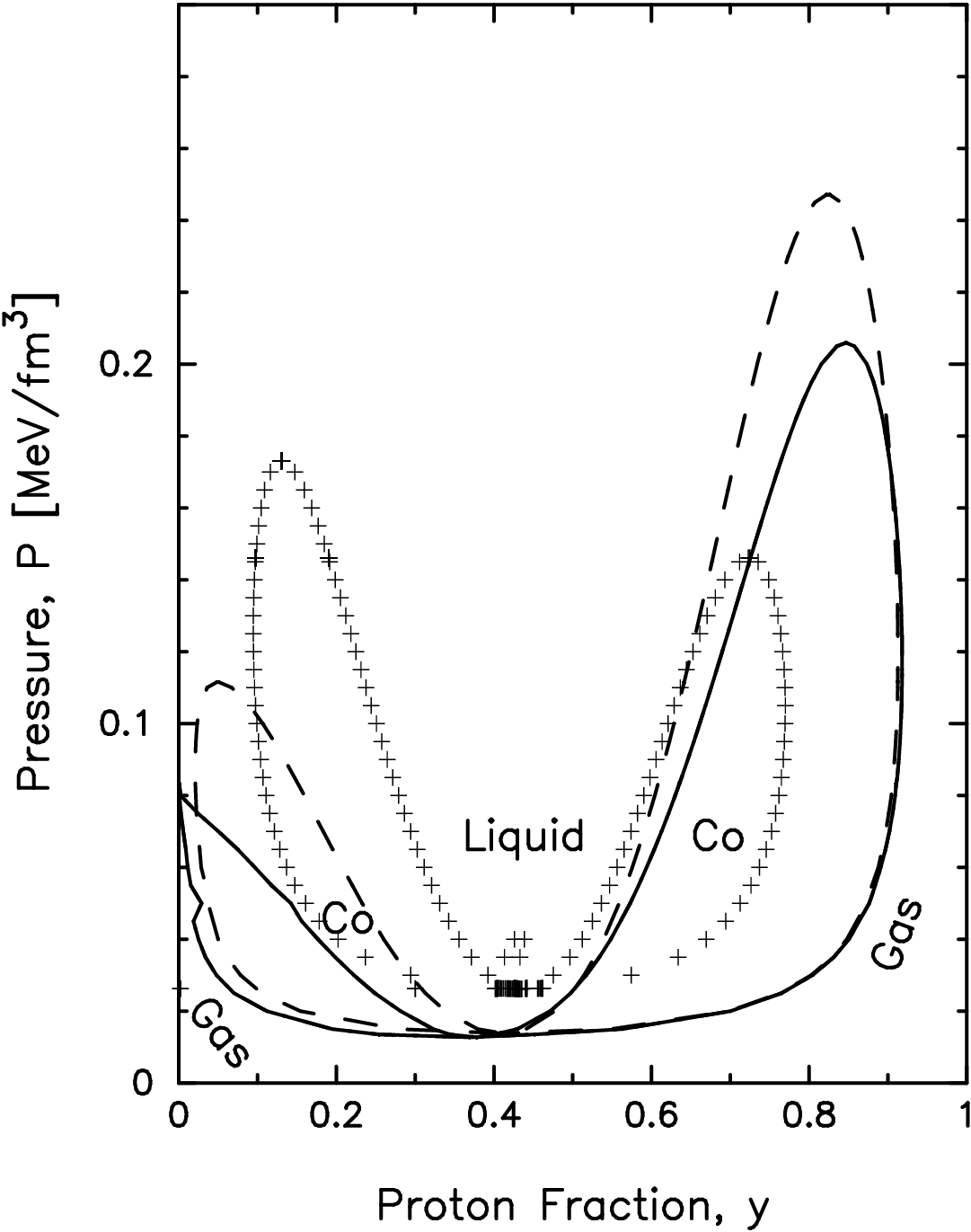}  
\caption{$P(y)$ plot for coexistence curve. 
The curves are the same as in Fig.1. 
The curve with + sign is for SLy4 parameter. }
     \label{fig5}
\end{figure}

\begin{figure}
\includegraphics[width=\figsiz]{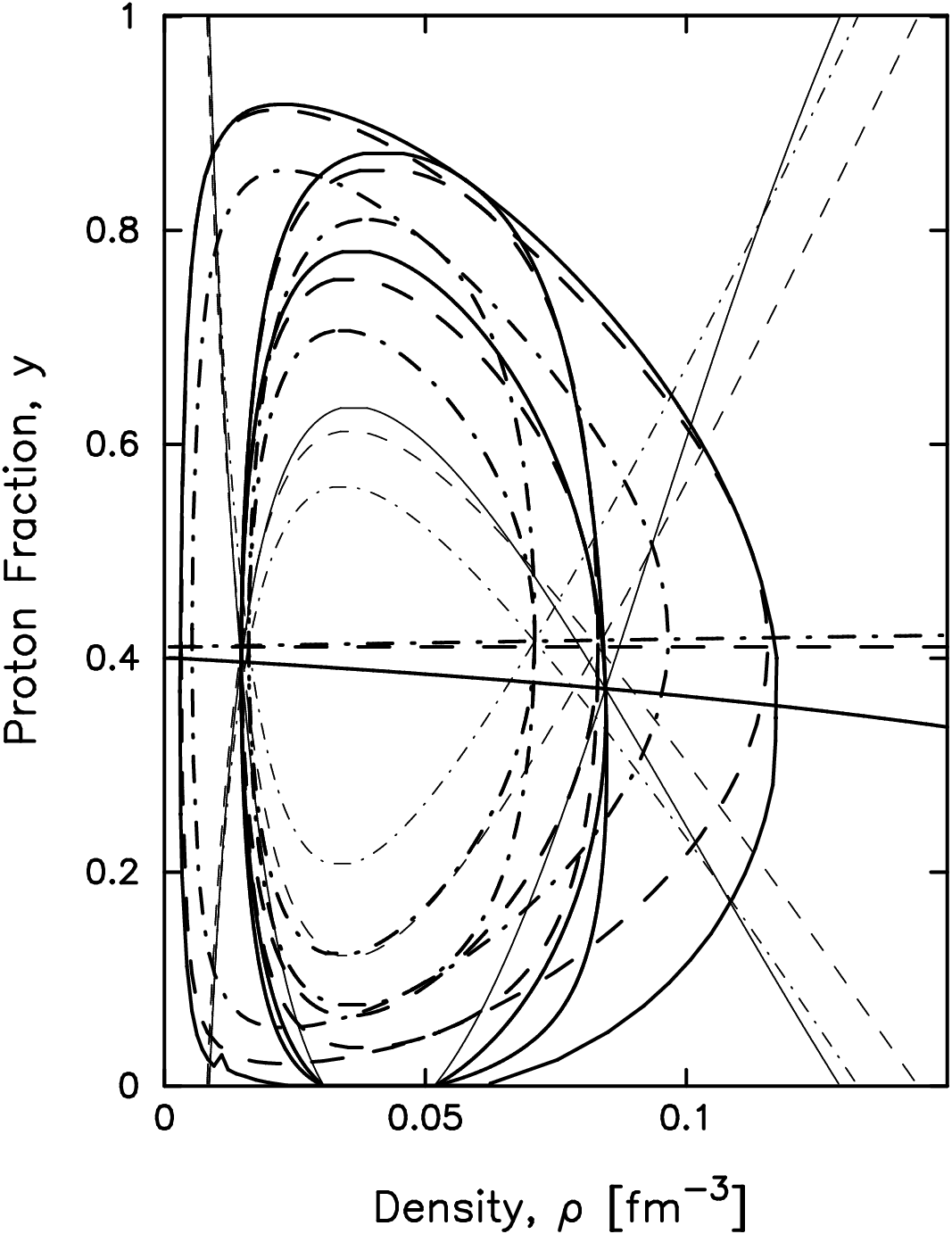}
\caption{The closed loops are for coexistence curves, chemical 
instability boundary curves, and mechanical instability curves 
from outmost loops to inside loops at $T = 10$ MeV.
Also shown by thin lines are $\partial \mu_q/\partial\rho = 0$
for proton (opened downward) and for neutron (opened upward). 
The horizontal lines are the $y_m(\rho)$ determine by Eq.(\ref{ymin}). }
     \label{fig6}
\end{figure}

For $y = 0$ and with an isovector momentum dependent term
the $P(\rho)$ curve has a spinodal instability region in the density 
range of about $1/5 < \rho/\rho_0 < 1/3$ of nuclear saturation density 
$\rho_0$ and around $P = 0.08$ (0.0755 $\sim$ 0.0834) MeV/fm$^3$.
In Figs.\ref{fig2} and \ref{fig4} the curves for $P = 0.08$ MeV/fm$^3$ show
a gap in the range of $0.0401 < \rho < 0.0598$ fm$^{-3}$ 
due to the spinodal behavior for $y = 0$. 
From solid curves of Fig.\ref{fig1} we can see the horizontal cut 
of $P = 0.08$ MeV/fm$^3$ gives negative value of $y$ in this density range.
This spinodal instability of $P(\rho)$ curve for $y = 0$ causes 
the cut of coexistence curve by the $y=0$ line 
which can be seen in Figs.\ref{fig5} and \ref{fig6}.

For $y = 1/2$ the $P(\rho)$ curve for momentum dependent interaction 
with both isoscalar and isovector terms (solid curve) overlaps 
with the curve for momentum independent interaction (dahed curve). 
The SKM($m^*=m$) interaction used here (Table \ref{tabl1}), 
with both isoscalar and isovector momentum dependent terms, 
has the effective mass of $m_q^* = m$ for $y = 1/2$ 
which is the same value with the momentum independent interaction case. 
The $y(\rho)$ curves at fixed $P$ in Fig.\ref{fig2} show that 
the curves for a momentum independent force (dashed line) are tangent 
to the curves for fully momentum dependent force (solid line) 
at $y = 1/2$. 
Similary the $\mu_q(y)$ curves in Fig.\ref{fig3} show that 
the dashed curve crosses the corresponding solid curve at $y = 1/2$. 
In Fig.\ref{fig4} this crossing appears at the point with large $\rho$ 
which corresponds to $y = 1/2$. 
Since $m^* = m$ at $y = 1/2$ for a full momentum dependent force 
the $P(\rho)$ curve at $y = y_m(\rho)$ is closer to the curve for 
momentum independent force than to the curve for a momentum dependent 
interaction without an isovector term.


Figs.\ref{fig2} and \ref{fig6} each show that the momentum dependent 
isovector term introduces a large $\rho$ dependence in the $y_m$ 
of Eq.(\ref{ymin}), the proton concentration $y$ having 
minimum pressure for a given density $\rho$.
The $y_m(\rho)$ is independent of the density $\rho$ for a momentum
independent Skryme force. 
The isoscalar momentum dependent term introduces a small increase
of $y_m$ with increasing $\rho$ while the isovector momentum dependent 
term introduces a large decrease of $y_m$ with increasing $\rho$.

Figs.\ref{fig1} - \ref{fig6} show that the basic behavior
of various curves remains qualitatively the same for the various 
cases considered here.
However the isoscalar momentum dependent term 
and the isovector momentum dependent term move in an 
opposite direction for the various curves compared to the curves 
of a momentum independent case. 
The isovector term has a larger effect than an isoscalar term. 

The $y(\rho)$ curve shown in Fig.\ref{fig2} is the horizontal cut 
of a fixed $P$ in Fig.\ref{fig1}.
Fig.\ref{fig2} shows the closed loops of $y(\rho)$ for low
pressure $P$ is reduced to a much smaller loop by the isoscalar 
momentum dependent term (dash-dotted line) from the momentum 
independent interaction (dashed line). 
The isovector momentum dependent term makes the loop much larger. 
Thus, the full momentum dependent interaction (solid line) enlarges 
the loop from the momentum independent loop. 
For higher $P$ which has no closed loop, the isoscalar momentum 
dependent term makes the density $\rho$ for a given $y$ and $P$ smaller 
than $\rho$ for momentum independent force 
and the isovector momentum dependent term makes $\rho$ larger. 
The full momentum dependent term makes $\rho$ for a given $y$ and $P$ 
larger than the $\rho$ for momentum independent force. 
The proton fraction $y$ for a given $\rho$ and $P$ is farther away 
from $y_m$ for full momentum dependent interaction than other interactions 
due to the smaller value of symmetry energy. 
The solid curve for $P = 0.08$ MeV/fm$^3$ shows a cut off of the curve 
by $y = 0$ line for the density range of 0.04 $\sim$ 0.06 fm$^{-3}$. 

Fig.\ref{fig3} and \ref{fig4} show that the chemical potential has a 
somewhat complicated relative effect arising from momentum dependent terms.
Except in the regions with medium $y$  
in Fig.\ref{fig3} or larger value of $\rho$ in Fig.\ref{fig4}, 
the isoscalar momentum dependent term makes the chemical potential 
higher while isovector momentum dependent term makes it lower. 
The closed loop for low $P$ in Fig.\ref{fig4} is reduced to a much 
smaller size by momentum dependent isoscalar term (dash-dotted curve) 
from a loop for a momentum independent interaction (dashed curve) 
while it is enlarged 
by momentum dependent isovector term (solid curve). 
The low $\mu_q$ portion of the curves for higher $P$ in Fig.\ref{fig4}, 
which corresponds to the  region with lower $y$ value, 
shows the density for a given $\mu_q$ becomming much smaller 
by the isoscalar momentum dependent term from the value for the momentum 
independent case while it becomes much larger by the isovector momentum 
dependent term. 
The horizontal straight thick solid line (the full momentum dependent 
SKM($m^*=m$) force) at low $\mu_p$ for $P = 0.08$ MeV/fm$^3$ in Fig.\ref{fig4} 
represents the cut due to the condition of $y \ge 0$ for the density range 
of $0.04 < \rho < 0.06$ fm$^{-3}$.

\subsection{Liquid-Gas Phase Transition and the Coexistence Curve 
and Instability}

For a one component system the coexistence curve is a line obtained by
the familiar Maxwell construction as already noted. For a two component 
system the coexistence region is a surface obtained as follows.
The condition for coexistence between the two phases requires the proton
chemical potentials to be the same in two phases and, similarly,
the neutron chemical potentials must be the same in the two phases at
a given pressure and temperature.
Note that the proton fraction or neutron fraction 
need not be the same in each of the two phases.
The two phases have the same proton fraction at the line of equal 
concentration $y_E$ which is the line of intersection between the 
coexistence surface and 
the surface of $y_m$ satisfying Eq.(\ref{ymin}). 
In fact, the liquid phase should be a more symmetric system than the gas 
phase because of the symmetry potential as seen in Refs.\cite{prc63,plb580}.
This observation goes under the name of isospin fractionation. 

Figs.\ref{fig5} and \ref{fig6} show features of the coexistence curves.
Also shown in Fig.\ref{fig6} are the mechanical and chemical instability 
loops and curves of $\partial\mu_q/\partial\rho = 0$. 
In Figs.\ref{fig5} and \ref{fig6} the small kink at lowest $y$ side of 
the coexistence loop for the interaction with an isovector momentum 
dependent term might have come from a numerical problem. 
For SLy4 (curve with + sign in Fig.\ref{fig5}), there is a small 
island of coexistence near the $y_E$ region.  
It is not clear if this is a numerical problem or 
a physical feature. Further detailed studies are required. 

Fig.\ref{fig5} shows that the coexistence loops in the neutron rich side 
(smaller $y$ side) are smaller than the loops in the proton rich side 
for three cases of SKM($m^*=m$) parameter sets. 
For the SLy4 parameter set, the coexistence loop at the neutron rich side 
is larger than the loop at the proton rich side. 
Ref.\cite{plb580} shows that the Coulomb interaction moves $y_E$ to a 
smaller proton fraction $y$ and makes the coexistence loops smaller 
in the neutron rich side than in the proton rich side. 
The SKM($m^*=m$) parameter sets used in Ref.\cite{plb580} has no isovector 
three body term ($x_3 = -1/2$) while SLy4 parameter sets has a negative 
three body isovector term. 

In Fig.\ref{fig6} we can see the instability curves are inside of 
the coexitence loop for all three force parameter sets. 
The mechanical instability loops are inside of the chemical instaility 
loops and tangent to each other at the point of $y_m$ lines (the horizontal 
line in this figure). 
The chemical instability loops are inside of the coexistence loops 
and tangent to each other at the critical point, i.e., the point 
with highest pressure on the coexistence curve. 

In Figs.\ref{fig5} and \ref{fig6}, the loops for the isovector momentum 
dependent case (solid line) has a cut at $y =0$ which is related with 
the spinodal behavior of $P(\rho)$ curve for $y = 0$ in Fig.\ref{fig1}. 
The smaller symmetry energy allows the system to be more asymmetric 
and thus to reach the boundary of $y = 0$ or $y = 1$. 
The SKM($m^*=m$) force with both isoscalar and isovector momentum dependent 
terms has the smallest symmetry energy while the SLy4 force has the 
largest symmetry energy among forces used here (see Table \ref{tabl1}). 
Fig.\ref{fig5} shows the coexistence loops for SLy4 are more symmetric 
than the loops for SKM($m^*=m$) parameter sets and the full SKM($m^*=m$) 
force has the most asymmetric coexistence loops. 
The cut of coexistence loop by $y = 0$ indicates that the SKM($m^*=m$) force 
with both isoscalar and isovector terms has too small of a symmetry 
energy for a description of nuclear system 
even though we have used $x_0 = -1/6$ instead of $x_0 = 0.5$ to have 
a larger symmetry energy. 

The pressure of the coexistence curve on the neutron rich side 
for a given value of $y$ (Fig.\ref{fig5}) becomes higher due to 
the momentum dependent isoscalar term (dash-dotted line) 
than the pressure for momentum independent interaction (dashed line) 
and also moves the peak to higher $y$.
By contrast the momentum dependent isovector term (solid line) makes 
the pressure lower and moves the peak to lower $y$.
The effects of isoscalar and isovector term work in opposite 
direction for the coexistence curve on the proton rich side compare
to the effects for neutron rich side. 
For SLy4 the system without an isovector or isoscalar momentum dependent 
term does not saturate at the usual normal density region. 
However, the isovetor momentum dependent term in SLy4 has the same 
sign as the isoscalar momentum dependent term 
thus both isoscalar and isovector terms would make the pressure of 
the coexistence curve of neutron rich side higher than momentum 
independent case. 
This might also make the coexistence loop in neutron rich side 
larger than the loop in proton rich side. 
In Fig.\ref{fig6} the loops for the momentum dependent isoscalar case 
(dash-dotted line) is smallest and thus mostly inside among the loops 
while the loops for the momentum dependent isovector case (solid line) 
are largest and thus mostly outside. 
The loops for the momentum independent case (dashed line) are tangent 
to the loops for the momentum dependent case with both isoscalar and 
isovector terms (solid line) at $y = 1/2$ where $m^* = m$ for both cases.
The curves of $\partial\mu_q/\partial\rho = 0$ (thin curves in Fig.\ref{fig6}) 
also show that the isovector momentum dependent term has an opposite and larger 
effect than the isoscalar momentum dependent term.

\section{Summary and Conclusions}

 A density functional theory based on a Skyrme nuclear 
interaction is used to study the temperature dependent 
properties of a two component system of strongly interacting 
protons and neutrons. 
The paper extends our previous studies by including an 
isovector momentum dependent interaction. 
As before, Coulomb and finite size surface effects are 
also contained in the description of various properties. 
The nuclear interaction has a velocity or 
momentum dependent term and in a medium this momentum dependence 
can be discussed as an effective mass to lowest order in it. 
The momentum dependence has both isoscalar and isovector terms. 
Our prior study in Ref.\cite{prc77} suppressed the isovector part 
by using a particular choice of Skyrme parameters. 
This greatly simplified the original calculations of 
the stability and coexistence properties. 
Here we now explore the mechanical and chemical instability of 
a two component system and also the coexistence features 
associated with a liquid gas phase transition incorporating 
an isovector momentum dependence besides other terms mentioned above. 
While qualitative results have a similar behavior with and 
without a momentum dependence (isoscalar and isovector), 
significant quantitative differences exist. 
Moreover, we extended our calculations to include the frequently 
used Skyrme interaction called SLy4 \cite{sly4}.

 Each term, Coulomb, volume, surface, symmetry, 
effective mass or momentum dependence, both isoscalar 
and isovector, play a unique role in determining 
the finite temperature stability properties and coexistence 
features as shown in various figures in this paper. 
For example, the Coulomb interaction leads to a 
pronounced asymmetry in the coexistence loops shown 
in Fig.\ref{fig5}. 
Moreover the equal concentration point $y_E$, 
where the liquid and gas have the same proton fraction,
moves away from $y_E = 1/2$ without Coulomb forces to a point 
where the proton fraction is close to the valley of stability 
when Coulomb forces are included. 
For example $y_E = 0.410$ for a momentum independent force, 
0.414 for a isoscalar momentum dependent force,
0.379 for a isovector momentum dependent force,
and 0.402 for SLy4. 
The $y_E$ with Coulomb included is somewhat insensitive to
the nuclear force. However, the pressure at $y_E$ is more 
sensitive. Specifically, $P = 0.0138$ MeV/fm$^3$ for momentum 
independent case, 0.0236 for isoscalar case, 0.0127 for isovector
case, and 0.0262 for SLy4.

A model with no Coulomb terms would have identical loops in $y$ 
around the point of intersection $y_E = 1/2$ of equal concentration. 
The SLy4 interaction has the interesting property of making 
the left coexistence loop of lower proton fraction more pronounced 
than the right loop of high proton fraction. 
This particular feature in the height of the right versus 
left loop is not observed with the other SKM($m^*=m$) interaction 
discussed in this paper. 
In particular, the right loop is still more pronounced or 
higher than the left loop for SKM($m^*=m$). 
For SKM($m^*=m$), the isoscalar momentum dependent term raises the pressure
of the coexistence loop on the neutron rich side while the isovector
momentum dependent term, which has an opposite sign from the isoscalar
term, lowers the pressure.
For SLy4, both the isoscalar and isovector terms have the same sign
and thus would raise the pressure of the coexistence loop on the neutron
rich side. The SLy4 has also a non zero effect from an isovector three 
body term which was zero for SKM($m^*=m$).
Besides this feature the SLy4 loop at higher $y$ side is narrower 
in $y$ for a given $T$ and $P$ than the SKM($m^*=m$) loops. 
From this observation we can conclude that the coexistence 
and stability behaviors are sensitive to the choice 
of Skyrme interaction parameters. 
The SLy4 has a higher symmetry energy than the other 
interactions used in this paper as shown in Table \ref{tabl1}. 
Moreover a higher left loop means that the liquid and 
gas phases can coexist at higher pressures at 
a given temperature with a more neutron rich gas and 
a lesser rich neutron liquid, 
i.e., $y_G < y_L < y_E < 1/2$ for the left loop. 
A narrowing of the coexistence loop brings the $y_G$ and $y_L$ 
closer together thereby reducing the proton fraction difference 
in the gas and liquid phases.

The momentum dependent isovector term in SKM($m^*=m$) force has an opposite 
and larger effect  than the momentum dependent isoscalar term. 
A momentum dependent isovector term makes the symmetry energy smaller 
and thus the coexistence loops and the mechanical and chemical 
instability loops become more asymmetric. 
The small value of symmetry energy for full momentum dependent 
interaction makes the $y$ value for a given $\rho$ and $P$ 
farther away from $y_m$. 
The pressure for a given $\rho$ and $y$ at fixed $T$ becomes larger 
by the isoscalar momentum dependent term while it becomes smaller by 
the isovector momentum dependent term.

Symmetry energy terms arising from momentum independent and 
momentum dependent terms tend to restore the symmetry of the loops. 
The coexistence loops and instability loops for SKM($m^*=m$) with both 
isoscalar and isovector momentum dependent terms have the smallest 
symmetry energy. Therefore they are the furthest away from $y_E$
for this force and thus have a cut due to the $y \ge 0$ condition.
This feature presents a problem for using an interaction with too
small of a symmetry term in neutron star studies.
The coexistence loops for SLy4, which has the largest symmetry 
energy, are closer to its $y_E$ than for the SKM($m^*=m$) parameter sets.
This suggests that the large symmetry energy terms
of the SLy4 are very important in restoring the isospin
symmetry in the system and also better in neutron star studies.

\acknowledgments
This work was supported in part 
by Grant No. KHU-20080646 of the Kyung Hee University Research Fund in 2008 
and by the US Department of Energy under DOE Grant No. DE-FG02-96ER-40987.

\end{document}